\documentclass[a4paper,11pt]{article}

\usepackage{jheppub} 
\usepackage{amsmath,amsfonts,amssymb}
\usepackage{mathtools}
\usepackage{ascmac}
\usepackage{physics}
\usepackage{color}

\def\bs#1{\boldsymbol{ #1 }} 
\newcommand{\wt}{\widetilde}

\newcommand{\ol}{\overline}
\newcommand{\del}{\partial}
\newcommand{\ra}{\rightarrow}

\newcommand{\nn}{\nonumber}

\numberwithin{equation}{section}

\arxivnumber{2602.21769} 

\title{\boldmath A Consistent Holographic Analysis of Anomaly-induced Charge Transport in the D3/D7 Model}
\author{Shin Nakamura}
\author{and Kensei Tanaka}%
\affiliation{%
 Department of Physics, Chuo University,\\ 1-13-27 Kasuga, Bunkyo-ku, Tokyo 112-8551, Japan}
\emailAdd{nshin001z@g.chuo-u.ac.jp}
\emailAdd{a20.p3h8@g.chuo-u.ac.jp}

\abstract{  

We propose a scheme to correctly incorporate the contribution of the chiral anomaly in the D3/D7 model to calculate chiral transport phenomena. To ensure the D7-brane wraps $S^5$ appropriately and the Wess-Zumino term is switched on, we allow the D7-brane to rotate in the compactified extra directions and perform the analysis accordingly. To demonstrate that this calculation procedure works well, we specifically compute the magnetoresistance in the D3/D7 model. We find that a finite axial chemical potential is realized and the negative magnetoresistance is enhanced by the anomaly contribution.

}

\begin{document}
\maketitle
\flushbottom

\section{Introduction}
Magnetoresistance refers to the change in electric resistance in response to an external magnetic field. In particular, when the electric resistance decreases with increasing magnetic field, the phenomenon is known as negative magnetoresistance.
There are various possible origins of negative magnetoresistance. Among them, negative magnetoresistance caused by the chiral anomaly \cite{Son:2012bg} has attracted significant attention.

The chiral anomaly is the phenomenon where the chiral symmetry is broken by the quantum effect \cite{PhysRev.177.2426,Bell:348417}. 
This phenomenon appears in high-energy physics,
and it explains, for example, the $\pi^0\to 2\gamma$ decay process \cite{PhysRev.177.2426,Bell:348417}.
The chiral anomaly is also important in condensed matter physics. In systems such as the Dirac semimetals or the Weyl semimetals, gapless fermions are effectively realized, and the system exhibits the chiral anomaly.
The negative magnetoresistance 
can be evidence of the existence of the chiral anomaly and is observed in the Dirac semimetals and Weyl semimetals (see, e.g. \cite{Li:2014bha,Huang:2015eia}).
\footnote{Note that negative magnetoresistance can arise from mechanisms unrelated to the chiral anomaly, and therefore careful verification is generally required \cite{Reis_2016,Ong:2020ffe}.}

In systems with the chiral anomaly, the axial charge density and the axial chemical potential are produced when an electric field and a magnetic field are applied in parallel to each other.\footnote{For a $1+1$-dimensional picture of this phenomenon, see \cite{Nielsen:1983rb}.} 
When a finite axial chemical potential and magnetic field exist, an electric current flows in the direction of the magnetic field as a chiral magnetic effect \cite{Nielsen:1983rb,PhysRevD.22.3080,Fukushima:2008xe}.
As a result, the electric resistance, which is the ratio between the electric field and the electric current, depends on the magnetic field, and the system exhibits negative magnetoresistance \cite{Son:2012bg}. 

In general, the axial charge dissipates, and a steady state is realized in which the decay and production rates of the axial charge are in balance. This steady state is a non-equilibrium steady state because of the presence of dissipation. 
The external electromagnetic fields continuously perform work on the system,
and energy is put into the system to maintain the steady state.

Although the chiral anomaly is topologically protected, the decay rate of the axial charge is not. Hence, the magnetoresistance depends on the details of the system. In particular, due to the presence of dissipation, the steady state is a non-equilibrium steady state, making the calculation of the magnetoresistance highly non-trivial. 
Holography, also referred to as the gauge/gravity duality or the AdS/CFT correspondence \cite{Maldacena:1997re,Gubser:1998bc,Witten:1998qj}, is one framework capable of handling non-equilibrium steady states in strongly interacting systems.

In the present paper, we consider magnetoresistance in the D3/D7 model of holography.
The nonlinear conductivity in the presence of magnetic field in this model was computed in \cite{Ammon:2009jt}, and a negative magnetoresistance was obtained \cite{Ammon:2009jt,Baumgartner:2017kme}.
However, since the contribution of the Wess-Zumino term representing the anomaly effect is absent, the negative magnetoresistance computed in \cite{Ammon:2009jt} is not due to the chiral anomaly \cite{Baumgartner:2017kme}.
On the other hand, since the mechanism of negative magnetoresistance from the axial charge production and the chiral magnetic effect is quite general, one would expect that the D3/D7 model should exhibit negative magnetoresistance originating from the chiral anomaly.

In this paper, we present computations of nonlinear conductivity in the presence of magnetic field incorporating the effects of the chiral anomaly in the D3/D7 model.
In particular, we show that the chiral anomaly in this system does contribute to the magnetoresistance, and we obtain the negative magnetoresistance that incorporates both the effects of the anomaly and effects unrelated to the chiral anomaly.

The reason why the anomaly effect was not included in previous calculations is that the Wess-Zumino term was not switched on due to insufficient winding of the D7-brane around the $S^5$ of the dual geometry.
In our computations, we relax the ansatz of the D7-brane configuration, and we employ the configuration winding the $S^5$ sufficiently: 
the D7-brane in our configuration is rotating in a subspace of the $S^5$. Actually, rotating D7-branes have been considered in \cite{OBannon:2008cmg,Das:2010yw,Hoyos:2011us,Guo:2016nnq}, and 
the axial chemical potential has been taken into account \cite{OBannon:2008cmg,Hoyos:2011us,Guo:2016nnq}.
We assert that the axial chemical potential is appropriately taken into account in our computations to obtain the negative magnetoresistance owing to the chiral  anomaly.\footnote{Magnetoresistance in the presence of chiral anomaly in the Sakai-Sugimoto model \cite{Sakai:2004cn,Sakai:2005yt} was studied in \cite{Fukushima:2021got}. In the Sakai-Sugimoto model, the D8-brane always wraps the $S^4$, and the Wess-Zumino term is switched on if the gauge field on the worldvolume adopts the appropriate configuration. However, it is necessary to extract a regulated part to examine the static longitudinal conductivity when the electric field and the magnetic field are not perpendicular to each other.}$^{,}$\footnote{For analysis  of negative magnetoresistance in other holographic models, see e.g. \cite{Landsteiner:2014vua,Jimenez-Alba:2014iia,Jimenez-Alba:2015awa,Sun:2016gpy}.}

The organization of the present paper is as follows. 
In Section \ref{sec:2}, we review the chiral transport phenomena. We revisit how a negative magnetoresistance is realized from the production of the axial charge and the chiral magnetic effect. 
In Section \ref{sec:3}, we introduce the D3/D7 model and review the previous computations of nonlinear conductivity, where a negative magnetoresistance is obtained even without the contribution of the chiral anomaly.
Section \ref{sec:4} is the main part of this paper. We present computations of magnetoresistance incorporating the contribution of the chiral anomaly. We relax the previous ansatz so that an appearance of finite axial chemical potential is allowed. 
We also perform numerical computations to show that the magnitude of negative magnetoresistance is enhanced in our computations by taking the contribution of the chiral anomaly appropriately.
We conclude in Section \ref{sec:5}.


\section{Chiral transport phenomena}\label{sec:2}
Let us briefly review chiral transport phenomena
arising from the chiral anomaly.\footnote{See for example, \cite{gorbar2021electronic,Landsteiner:2016led}. }
We consider a system of massless Dirac fermions that couple to an external electromagnetic field.
When we apply an electric field $\bs{E}$ and a magnetic field $\bs{B}$ in such a way that $\bs{E}\cdot\bs{B}\neq 0$, the conservation law of the axial current $j^\mu _5$ is broken by the chiral anomaly \cite{PhysRev.177.2426,Bell:348417}:
\begin{equation}\label{eq:anomaly}
  \del _\mu j^\mu _5 = \frac{1}{2\pi^2} \bs{E}\cdot \bs{B}.
\end{equation}

First, we review how the axial charge density, which we denote by $n_5$, is produced by the chiral anomaly.
Let us consider the cases where the system is spatially homogeneous.
Then the left-hand side of (\ref{eq:anomaly}) is $\partial_{t}n_{5}$,
and this indicates that the axial charge density is produced by $\bs{E}\cdot \bs{B}$.

When a fermion is massive, the chiral symmetry is explicitly broken, and we have a mass-dependent correction term in (\ref{eq:anomaly}).
In general, 
if interactions allow chirality flipping,
(\ref{eq:anomaly}) may be effectively modified as follows:
\begin{equation}
\partial_{t}{n}_5  =  \frac{1}{2\pi^2} \bs{E} \cdot \bs{B} -\frac{n_5}{\tau},
\end{equation}
where $\tau$ is the relaxation time.
We assume that the electromagnetic field does not depend on time. Suppose that sufficient time has elapsed since a static electromagnetic field was switched on, and the system has reached a steady state. Then $\partial_{t}{n}_5=0$ and we obtain
\begin{equation}\label{eq:n5}
  n_5=\frac{\tau}{2\pi^2} \bs{E} \cdot \bs{B},
\end{equation}
which indicates that the system acquires a finite axial charge density.
Furthermore, considering that a relationship $n_5=\mu_5 \chi$ between the axial charge density $n_5$ and the axial chemical potential $\mu_5$ with the susceptibility $\chi$, the system achieves a nonzero axial chemical potential in the steady state.

Another physical effect arising from the chiral anomaly is the chiral magnetic effect.
The chiral magnetic effect is a phenomenon where a vector current that is parallel to a magnetic field arises owing to the chiral anomaly without any electric field \cite{PhysRevD.22.3080,Fukushima:2008xe}.
In $3+1$ dimensional systems of Dirac fermions, the chiral magnetic current $\bs{j}_{\mathrm{CME}}$ is given by
\begin{eqnarray}
  \bs{j}_{\mathrm{CME}}=\frac{\mu_5}{2 \pi ^2}\bs{B} \label{eq:CME} .
\end{eqnarray}
Combining the two effects mentioned above, we obtain
\begin{eqnarray}
  \bs{j}_{\mathrm{CME}}=\frac{\bs{E} \cdot \bs{B}}{(2 \pi ^2)^2\chi}\bs{B} \label{eq:CME-2}.    
\end{eqnarray}
When the electric field and the magnetic field are parallel, the electric conductivity is proportional to $B^2$ if the susceptibility is a constant.

In general, we may have a current that is not related to the chiral anomaly. If we express such current densities as $\bs{j}_{\mathrm{Ohm}}=\sigma _{\mathrm{Ohm}}\bs{E}$, the total current density $\bs{j}_{\mathrm{tot}}$ is given by
\begin{equation}
  \bs{j}_{\mathrm{tot}} =\left( \frac{\tau}{(2\pi ^2)^2 \chi }B^2 + \sigma _{\mathrm{Ohm}}\right)\bs{E}.
\end{equation}
The electric resistivity $\rho$ for this case is given by
\begin{equation}
  \rho = \left[ \frac{\tau}{(2\pi ^2)^2 \chi }B^2 + \sigma _{\mathrm{Ohm}} \right]^{-1},
\end{equation}
which indicates a negative magnetoresistance that arises from the chiral anomaly.


\section{Previous computations in the D3/D7 model}\label{sec:3}
\subsection{The D3/D7 model}\label{subsec:31}
To analyze strongly coupled systems of fermions, we consider the D3/D7 model, a holographic model based on a system composed of $N_c$ D3-branes and one D7-brane \cite{Karch:2002sh}. (We set the number of flavors to $1$.)

First, let us summarize the field theory that is realized on the D3/D7 brane systems on a flat $9+1$-dimensional spacetime.
Table \ref{tab:D3D7} shows the directions in which these D-branes are extended.
\begin{table}[h]
  \centering
  \caption{$\rm{D3 /D7}$-brane configuration
  }\label{tab:D3D7}
  \begin{tabular}{|c||c|c|c|c|c|c|c|c|c|c|}
      \hline
       & $X^0$ & $X^1$  & $X^2$ & $X^3$ & $X^4$ & $X^5$ & $X^6$ & $X^7$ & $X^8$& $X^9$\\ \hline
      D3 & \checkmark  & \checkmark  & \checkmark  & \checkmark  &  & &  &  & &  \\ \hline
      D7 & \checkmark  & \checkmark  & \checkmark  & \checkmark  & \checkmark  & \checkmark  & \checkmark & \checkmark  &  &   \\ \hline
  \end{tabular}
\end{table}
A $U(N_c)$ $\mathcal{N}=4$ Super Yang-Mills (SYM) theory is realized on the D3-brane. 
We have an $\mathcal{N}=2$ hypermultiplet arising from the open strings between the D3-branes and the D7-brane. The hypermultiplet contains two Weyl fermions and two complex scalar fields. In this paper, we refer to the hypermultiplet sector as the flavor sector.

The particles of the $\mathcal{N}=4$ SYM are in the adjoint representation, whereas those in the flavor sector are in the fundamental or anti-fundamental representation.
Thus the degrees of freedom of the $\mathcal{N}=4$ SYM are $O(N_c^2)$, and those of the flavor sector are $O(N_c)$.\footnote{We consider only the cases where the $\mathcal{N}=4$ SYM is in the deconfinement phase, in this paper.}
Therefore, in the limit of $N_c \gg 1$, the $\mathcal{N}=4$ SYM sector possesses significantly greater degrees of freedom than the flavor sector and behaves as a heat bath at finite temperature.

The vacuum state of the flavor sector has a global $SO(4)\times U(1)_R$ symmetry.
The $U(1)_R$ transformation is realized as a rotation transformation within the $X^8$-$X^9$ plane in the D3/D7 system, and is interpreted as a $U(1)$ axial transformation ($U(1)_A$ transformation) for the fermions in the flavor sector. 
This will be revisited in Section \ref{subsec:41}. 

Let us consider the gravity dual of this system.
We consider the case where $N_c \gg 1$ and the 't Hooft coupling $\lambda = g_{YM}^2 N_c \gg 1$, where the gravity dual is described by using a probe D7-brane with the classical theory of type IIB supergravity.
Note that the gauge group of the corresponding field theory is now $SU(N_c)$, since the $U(1)$ part of the $U(N_c)$ gauge symmetry is decoupled. 

Unless otherwise specified, we deal with systems at finite temperatures in this paper. We consider only the case where the boundary is planar, hence the system is always in the deconfinement phase.
Then, the geometry of the gravity dual is given by
\begin{gather}
  \dd s^{2}=\frac{L^2}{u^{2}}\left(-\frac{f(u)^{2}}{h(u)}\dd t^{2}+h(u)\dd \vec{x}^{2} +\dd u^2 \right) +  L^2\dd {\Omega_5}^{2},\\
  f(u) = 1 - \left(\frac{u}{u_H}\right)^4 ,\quad h(u) = 1 + \left(\frac{u}{u_H}\right)^4  .
\end{gather}
This spacetime is composed of a $4+1$-dimensional AdS-Schwarzschild spacetime and an $S^5$.
Here, $(t, \vec{x})$ are the $3+1$-dimensional coordinates on which the gauge theory resides, and $u$ represents the radial direction in the AdS-Schwarzschild spacetime.
The horizon is located at $u=u_H$ and the boundary is at $u=0$.
The Hawking temperature, which corresponds to the temperature of the gauge theory, is given by
\begin{equation}
  T = \frac{\sqrt{2}}{\pi u_H }.
\end{equation}
$\dd \Omega_{5}$ is the line element of unit $S^5$ and is given as
\begin{gather}
\dd \Omega_{5}^2=\dd \theta^{2}+\sin^{2}\theta \dd \Psi^{2}+\cos^{2}\theta \dd \Omega_{3}^{2},
\end{gather}
where $\dd \Omega_{3}$ is the line element of unit $S^3$.
The D7-brane wraps a $S^3$ part of the $S^5$, and $L \cos\theta \dd \Omega_{3}$ represents the line element of the $S^3$ part.

There is a nontrivial background Ramond-Ramond 4-form potential
\begin{gather}
  C_{4}=\frac{L^4}{u^{4}}\dd t\wedge \dd x\wedge \dd y\wedge \dd z-L^4 \cos^{4}\theta \dd \Psi \wedge\omega(S^{3}), \label{eq:RR}
\end{gather}
where $\omega(S^3)$ is the volume element of the unit $S^3$. 

The action of the D7-brane, which we denote as $S_{\text{D7}}$, consists of the sum of the Dirac-Born-Infeld (DBI) action $S_{\mathrm{DBI}}$ and the Wess-Zumino (WZ) term $S_{\mathrm{WZ}}$:

\begin{gather}\label{eq:D3D7Action}
    S_{\text{D7}}=S_{\mathrm{DBI}} +  S_{\mathrm{WZ}}, \\ 
  S_{\mathrm{DBI}}=-T_{\mathrm{D7}}\int \dd ^{8}\xi\sqrt{-\det(g_{ab} +2\pi\alpha^{\prime}F _{ab})},\label{DBI}\\
  S_{\mathrm{WZ}} =\frac{(2\pi\alpha^{\prime})^{2}}{2}T_{\text{D7}}\int P[C_{4}]\wedge F\wedge F,
\end{gather}
where $T_\mathrm{D7}=(2\pi )^{-7}g_\mathrm{s}^{-1}l_\mathrm{s}^{-8}$ is the tension of the D7-brane and $\xi^a (a=0,1,2,\cdots ,7)$ are its worldvolume coordinates.
$g_{ab}$ is the induced metric of the D7-brane, and $P[C_4]$ is the pullback of the 4-form field. 
$F_{ab}$ is the field strength of the worldvolume $U(1)$ gauge field $A_{a}$ and $F = \dd A$.
We employ the static gauge:
the worldvolume coordinates are $t,\vec{x},u$ and those on the $S^3$.
The configuration of the D7-brane is specified by $\theta$ and $\Psi$ as functions of the worldvolume coordinates.
From now on, we set $(2\pi \alpha')=1$ for simplicity.

The mass of the hypermultiplet in the D3/D7 model is given by the distance between the D3-brane and D7-brane at the boundary.
Specifically, $\theta(u)$ is expanded with respect to $u$ near the boundary as
\begin{equation}
  \theta (u)= m u + c_3 u^3 + \cdots,
  \label{eq:mass}
\end{equation}
where $m$ represents the mass of the hypermultiplet,
and $c_3$ gives the chiral condensate up to appropriate normalizations (see \cite{Kobayashi:2006sb} for the exact normalization and corresponding operator).

\subsection{Negative magnetoresistance without anomaly}\label{subsec:32}
Calculations of nonlinear electric conductivity in the D3/D7 model in the presence of a magnetic field are presented in \cite{Ammon:2009jt}. 
Interestingly, this system exhibits negative magnetoresistance
even without including the chiral anomaly contribution \cite{Ammon:2009jt,Baumgartner:2017kme}.
Here, for reference in later sections, we review the electric conductivity in the D3/D7 model when electric and magnetic fields are applied parallel to each other \cite{Karch:2007pd, Ammon:2009jt}.

We employ the following ansatz to impose an electric field $E_x$ and a magnetic field $B_x$ in the $x$-direction:
\begin{equation}
  \begin{aligned}
    A_x (t,u) &= - E_x t + a_x (u) ,\\
    A_y (u) &=  a_y (u),\\
    A_z (y,u) &= B_x y + a_z (u) .
  \end{aligned}
    \label{ansatzA}
\end{equation}
For $\theta$ and $\Psi$,
we employ the following ansatz,
\begin{eqnarray}
  \theta &=& \theta(u) ,\label{ansatztheta}\\
  \qquad \Psi &=& \phi_0 , \label{constPsi}
\end{eqnarray}
where $\phi_0$ is a constant.
Note that the configuration $\Psi = \phi_0$ prevents the Wess-Zumino term from being switched on, and the chiral anomaly does not contribute to the conductivity.
In fact, $\Psi = \phi_0$ is not a sufficiently general ansatz. As we will see later, relaxing this ansatz to a more general one allows the chiral anomaly effect to be correctly incorporated. However, we review the results obtained using $\Psi = \phi_0$ in this subsection.

We obtain the following D7-brane action per unit volume of the $3+1$-dimensional spacetime:
\begin{equation}\label{eq:AmmonAction}
  \begin{gathered}
    S_{\text{D7}}
    =-\mathcal{N} \int \dd u \sqrt{w_1(u) + w_2(u) a_x'(u)^2 + w_3(u) \left(a_y'(u)^2 + a_z'(u)^2\right)},
  \end{gathered}
\end{equation}
where $\mathcal{N}= T_{\mathrm{D7}} (2\pi^2)$ and 
\begin{equation}
  \begin{aligned}
  w_1(u) &= -L^6 \cos ^6\theta \left({E_x}^2 + g_{tt} g_{xx}\right) \left({B_x}^2 + {g_{xx}}^2\right)  \left(g_{uu} + g_{\theta\theta} \theta'(u)^2\right), \\
  w_2(u) &= -L^6 \cos ^6\theta \, g_{tt} \left({B_x}^2 + {g_{xx}}^2\right) , \\
  w_3(u) &= -L^6 \cos ^6\theta\, g_{xx} \left({E_x}^2 + g_{tt} g_{xx}\right) .
\end{aligned}
\end{equation}
Since the action depends on $a'_x(u)$, $a'_y(u)$ and $a'_z(u)$ but not on $a_x(u)$, $a_y(u)$ and $a_z(u)$, the following quantities are constants: 
\begin{align}
   -\frac{1}{\mathcal{N}}\frac{\partial \mathcal{L}_{\mathrm{D7}}}{\partial a_x'(u)} &= \frac{ w_2(u) a_x'(u)}{\sqrt{w_1(u) + w_2(u) a_x'(u)^2 + w_3(u) \left(a_y'(u)^2 + a_z'(u)^2\right)}} \equiv C_x  \nn,\\
  -\frac{1}{\mathcal{N}}\frac{\partial \mathcal{L}_{\mathrm{D7}}}{\partial a_y'(u)} &= \frac{  w_3(u) a_y'(u)}{\sqrt{w_1(u) + w_2(u) a_x'(u)^2 + w_3(u) \left(a_y'(u)^2 + a_z'(u)^2\right)}}\equiv C_y \nn,\\
  -\frac{1}{\mathcal{N}}\frac{\partial \mathcal{L}_{\mathrm{D7}}}{\partial a_z'(u)} &=\frac{  w_3(u) a_z'(u)}{\sqrt{w_1(u) + w_2(u) a_x'(u)^2 + w_3(u) \left(a_y'(u)^2 + a_z'(u)^2\right)}} \equiv C_z,
\end{align}
where $\mathcal{L}_\mathrm{D7}$ is the Lagrangian density.
Given these constants, we find that the gauge fields behave at the boundary as
\begin{equation}
  \begin{aligned}
    a_x(u)&= (\text{const.}) + \frac{C_x}{2}u^2 + \mathcal{O} (u^3),\\
    a_y(u)&=(\text{const.})+ \frac{C_y}{2}u^2 + \mathcal{O}(u^3),\\
    a_z(u)&=(\text{const.}) + \frac{C_z}{2}u^2 + \mathcal{O} (u^3).
  \end{aligned}
\end{equation}
Using the GKP-Witten prescription, we identify
\begin{equation}
    C_x =\frac{j_x}{\mathcal{N}}  , \qquad C_y = \frac{j_y}{\mathcal{N}}  , \qquad C_z = \frac{j_z}{\mathcal{N}} ,
\end{equation}
where $j_i$ with $i=x, y, z$ gives the expectation value of the electric current density along the $i$-direction.

Let us perform the following Legendre transformation:
\begin{align}
  \wt{\mathcal{L}}_{\text{D7}} &= \mathcal{L}_{\text{D7}}  -  \left(a_x'(u) \frac{\partial \mathcal{L}_{\mathrm{D7}}}{\partial a_x'(u)} + a_y'(u) \frac{\partial \mathcal{L}_{\mathrm{D7}}}{\partial a_y'(u)} +a_z'(u) \frac{\partial \mathcal{L}_{\mathrm{D7}}}{\partial a_z'(u)} \right), \nn \\
  &=\mathcal{N}\sqrt{  w_1(u) - \frac{w_1(u)}{w_3(u)}(C_y^2 + C_z^2) - \frac{w_1(u)}{w_2(u)}C_x^2}.
  \label{LegendreL}
\end{align}
It is useful to present the following results:
\begin{equation}
  \begin{aligned}
  w_1(u)=&-L^6 \cos ^6\theta \left({E_x}^2 - |g_{tt}| g_{xx}\right) \left({B_x}^2 + {g_{xx}}^2\right)  \left(g_{uu} + g_{\theta\theta} \theta'(u)^2 \right),\\
  \frac{w_1(u)}{w_3(u)} =&\frac{\left({B_x}^2 + {g_{xx}}^2\right) \left(g_{uu} + g_{\theta\theta} \theta'(u)^2\right)}{g_{xx}}  > 0,\\
  \frac{w_1(u)}{w_2(u)}=&-\frac{\left({E_x}^2 - |g_{tt}| g_{xx}\right) \left(g_{uu} + g_{\theta\theta} \theta'(u)^2\right)}{|g_{tt}|} .
\end{aligned}
\end{equation}
Let us define $u_\ast$ using the following equation:
\begin{equation}
  \left. \left({E_x}^2 - |g_{tt}| g_{xx}\right)\right|_{u= u_\ast}=0.
\end{equation}
We call the location $u=u_\ast$ as the effective horizon.
When $u=u_\ast$, the quantity inside the square root in (\ref{LegendreL}) is negative unless $C_y=C_z=0$: we choose $C_y=C_z=0$ for the reality of $\wt{\mathcal{L}}_{\text{D7}}$.
Then we obtain
\begin{align}
  \wt{\mathcal{L}}_{\text{D7}} =\mathcal{N}\sqrt{ w_1(u)\left(1 - \frac{C_x^2}{w_2(u)}\right) }.
\end{align}
Note that $w_1(u)$ is positive at the horizon $u=u_H$ and negative at the boundary $u=0$, and $w_1(u)$ flips the sign at $u_\ast$.
In order to make $\wt{\mathcal{L}}_{\text{D7}}$
to be real for all values of $u$, 
$1 - C_x^2/w_2(u)$ must also change the sign at $u_\ast$ and therefore is vanishing there.
From this condition, $j_x$ is given by
\begin{align}
  j_x 
  &= \mathcal{N} E_x \sqrt{g_{xx}(u_\ast )^{-1}(B_x^2 + g_{xx}(u_\ast )^2)L^6 \cos ^6 (\theta(u_\ast ))} ,\nn \\
  &=\frac{\mathcal{N} E_x L^3 \cos^3(\theta(u_\ast)) \sqrt{4 L^4 + \left({B_x}^2 + {E_x}^2\right) {u_H}^4} }{u_H \left(4 L^4 + {E_x}^2 {u_H}^4\right)^{\frac{1}{4}}},
  \label{jx_ammon}
\end{align}
where the $\theta(u_\ast)$ should be obtained by solving the equation of motion for $\theta(u)$ that depends on $E_x$ and $B_x$.
Then the resistivity $\rho$ is obtained by
\begin{equation}\label{eq:rhoAmmon}
    \rho = \frac{u_H \left(4 L^4 + {E_x}^2 {u_H}^4\right)^{\frac{1}{4}}}{\mathcal{N} L^3 \cos^3(\theta(u_\ast)) \sqrt{4 L^4 + \left({B_x}^2 + {E_x}^2\right) {u_H}^4}}.
\end{equation}
We need numerical computation to obtain $\rho$ precisely. One finds that (\ref{eq:rhoAmmon}) exhibits negative magnetoresistance. This is explicitly shown in Figure \ref{fig:PLOT2}.
Therefore, the present setup yields negative magnetoresistance even without the contribution of the anomaly \cite{Baumgartner:2017kme}.

In our setup, where the electric and magnetic fields are parallel to each other, the right-hand side of (\ref{eq:anomaly}) has to be turned on. Then it is natural that the contribution of the anomaly appears in the conductivity.
In the following section, we will relax the ansatz (\ref{constPsi}) to a more general one to properly incorporate the contributions from the anomaly.
We will introduce the axial chemical potential and analyze the magnetoresistance with the anomaly effects properly incorporated.


\section{Negative magnetoresistance from anomaly}\label{sec:4}
\subsection{Introduction of axial chemical potential}\label{subsec:41}
In this section, we review how the axial chemical potential is realized in the D3/D7 systems \cite{Hoyos:2011us}.
First, let us consider the case where the D7-brane lies at $X^8=X^9=0$.
In this case, the system is invariant under the rotation within the $X^8$-$X^9$ plane;
\begin{equation}\label{eq:rotation}
  X^8+iX^9\to e^{i\phi}(X^8+iX^9).
\end{equation}
This symmetry corresponds to the $U(1)_R$ symmetry, which is a part of the R-symmetry of the hypermultiplet.

The right-handed (left-handed) Weyl fermions in the hypermultiplet carry R-charge of \(+1/2\) (\(-1/2\)) with respect to that of the complex scalar field \(X^8 + i X^9\).
Therefore, the Dirac fermion \(\psi\) constructed from these Weyl fermions transforms as
\begin{equation}
  \psi \ra e^{i \gamma ^5 \phi /2} \psi,
  \label{eq:U1axial}
\end{equation}
under the \(U(1)_{R}\) transformation (\ref{eq:rotation}). (\ref{eq:U1axial}) is also understood as the \(U(1)\) axial (\(U(1)_A\)) symmetry.
The chiral transformation we discuss in this paper is the $U(1)_A$ transformation. 
When the hypermultiplet are massive, the $U(1)_A$ symmetry is explicitly broken, but the correspondence between the rotation in the $X^8$-$X^9$ plane and the $U(1)_A$ transformation still holds.

Let us consider the case where the Dirac fermion $\psi$ is massive and focus on its kinetic and mass terms.
The kinetic and mass terms of $\psi$ transform as follows under transformations (\ref{eq:rotation}) and (\ref{eq:U1axial})
\begin{equation}
  \ol{\psi}(i\gamma^\mu \partial_\mu - m)\psi
  \to
    \ol{\psi}i\gamma^\mu \partial_\mu\psi
  +\frac{\del_\mu \phi}{2} \ol{\psi}\gamma^\mu \gamma^5 \psi
 -m\ol{\psi}e^{i\gamma^5 \phi}\psi.
\end{equation}
Therefore, we see that $\frac{1}{2}\del_\mu \phi$ and $\ol{\psi}\gamma^\mu \gamma^5 \psi=j^{\mu}_5$ are conjugate to each other.

Let us consider the case where $\phi=\omega t$ \cite{Das:2010yw,Hoyos:2011us}.
In this case, since $\frac{1}{2}\omega$ and $n^{5}$ are conjugate, $\frac{1}{2}\omega$ plays the role of the axial chemical potential $\mu_5$.
On the gravity side, this is realized by a D7-brane rotating in the $X^8$-$X^9$ plane at angular velocity $\omega$.
It should be noted that rotating the D7-brane in this manner generates not only an axial chemical potential, $\mu_5 = \omega/2$, but also an axially rotating Dirac mass term of the form $m\bar{\psi} e^{i\gamma^5 \omega t}\psi$ \cite{Das:2010yw}.

Now, let us consider the GKP-Witten prescription related to $\phi$ on the gravity side.
Suppose that the configuration $\Psi(x^{\mu},u)$ of the D7-brane is given by the following expression:
\begin{equation}
   \Psi(x^{\mu},u)= \phi(x^{\mu})+\Phi(u).
\end{equation}
Here, $\Phi(u)$ is assumed to vanish at the boundary, and $\phi(x^{\mu})$ is the boundary value of $\Psi(x^{\mu},u)$.
Since $\frac{1}{2}\del_\mu \phi$ and $\ol{\psi}\gamma^\mu \gamma^5 \psi=j_5^{\mu}$ are conjugate to each other, the expectation value of the axial current is given by the on-shell action $\mathcal{L}$ of the D7-brane as
\begin{equation}  \label{eq:j5mu}
   j_5^\mu  
  =2 \int \dd u \, \pdv{\mathcal{L}}{ (\del_\mu \phi)}
  =2 \int \dd u \, \pdv{\mathcal{L}}{ (\del_\mu \Psi)},\qquad (\mu=t,x,y,z).
\end{equation}
As we will explicitly see later, $\mathcal{L}_{\mathrm{D7}}$ does not explicitly contain $\Psi$,
and the equation of motion for $\Psi$ is
\begin{equation}\label{eq:eomPsi}
  \del_\mu \left( \frac{\del \mathcal{L}}{\del (\del_\mu \Psi)}\right)+\del_u \left( \frac{\del \mathcal{L}}{\del (\del_u \Psi)}\right) =0.
\end{equation}
Using (\ref{eq:j5mu}) and (\ref{eq:eomPsi}), we obtain the following equation:
\begin{align}\label{eq:divj5mu}
   \del_{\mu}j_5^\mu  
  &=-2 \int \dd u \, \del_u \left( \frac{\del \mathcal{L}}{\del (\del_u \Psi)}\right) \nonumber \\
  &=2\left( \left. \frac{\del \mathcal{L}}{\del (\del_u \Psi)}\right|_{u=0} - \left. \frac{\del \mathcal{L}}{\del (\del_u \Psi)}\right|_{u=u_\mathrm{IR}}\right)  .
\end{align}
Assuming that the spatial derivative of $j_5^\mu$ vanishes, the left-hand side of (\ref{eq:divj5mu}) represents the time evolution of the axial charge density, and the right-hand side reflects its generation and dissipation.
For the steady state, the right-hand side should vanish as discussed in Section \ref{sec:2}.

Here, $u=u_\mathrm{IR}$ is the IR boundary of the domain of integrations in (\ref{eq:j5mu}) and that in (\ref{eq:divj5mu}).
The contribution of $\frac{2\del \mathcal{L}}{\del (\del_u \Psi)}$ at $u=u_\mathrm{IR}$ in (\ref{eq:divj5mu}) 
represents the amount of the flow of axial charge into the region of $u>u_\mathrm{IR}$ in the bulk, and thus the magnitude of the dissipation of the axial charge into the thermal bath in the field-theory perspective \cite{Hoyos:2011us,Guo:2016nnq}.\footnote{
In our systems, the axial charge dissipates into the thermal bath when $\mu_5 \neq 0$.
In this case, the flavor sector is in a non-equilibrium steady state, and an effective horizon is formed on the D7-brane's worldvolume at $u = u_{*}$ \cite{Das:2010yw}. 
Although it is open to debate where to set $u_\mathrm{IR}$ (i.e., whether to set it to $u_{H}$ or $u_{*}$), we will not address this issue in this paper.
In any case, the choice of $u_\mathrm{IR}$ does not affect the subsequent discussions in this paper, since $\frac{\del \mathcal{L}}{\del (\del_u \Psi)}$ is a conserved quantity independent of $u$.

}
On the other hand, the contribution of $\frac{2\del \mathcal{L}}{\del (\del_u \Psi)}$ at $u=0$ in (\ref{eq:divj5mu}) is the system's response to the variation $\phi \to \phi + \delta \phi$, and represents the $U(1)_A$ symmetry breaking that includes the explicit breaking by the mass term and the breaking effect arising from the chiral anomaly.

It is interesting to note that the dissipation of the axial charge occurs due to the friction caused by the D7-brane being dragged in the $\phi$-direction, from the perspective of the higher-dimensional spacetime on the gravity side.
Although the dissipation of the axial charge and that of the momentum are distinct phenomena from the perspective of field theory, it is intriguing that they can be understood from a common viewpoint within the framework of the gauge/gravity duality.

\subsection{Nonlinear conductivity and anomaly}\label{subsec:42}
In light of the discussion in the previous subsection, we adopt the following ansatz instead of (\ref{constPsi}): 
\begin{equation}
 \Psi = \omega t + \Phi (u),
 \label{PsiAnsatz}
\end{equation}
to take the axial chemical potential into account. 
In this ansatz, we assume homogeneity in the spacial directions. 
We further assume that the internal $SO(4)$ symmetry is preserved and that the D-brane configuration is homogeneous along the $S^3$ directions.
This ansatz was introduced in \cite{Hoyos:2011us}. Another ansatz where $\phi(x^\mu)$ depends on a spatial coordinate was employed in \cite{BitaghsirFadafan:2020lkh,Kharzeev:2011rw} for a different aim.
We employ the same ansatz (\ref{ansatzA}) for the $U(1)$ gauge fields and that (\ref{ansatztheta}) for $\theta$.
The D7-brane action (\ref{eq:D3D7Action}) is now given by
\begin{align}
  S_{\mathrm{DBI}} &= - \mathcal{N} \int \dd u \, \bigg( w_1(u) + w_2(u) a_x'(u)^2 + w_3(u) \left( a_y'(u)^2 + a_z'(u)^2 \right), \nonumber \\
  &\qquad \qquad + w_4(u) \Phi'(u)^2 + w_5(u) a_x'(u) \Phi'(u) \bigg)^{\frac{1}{2}}, \\
  S_{\mathrm{WZ}} &= \mathcal{N} \int \dd u \, w_6(u) \left( \omega a_x'(u) + E_x \Phi'(u) \right),
\end{align}
where
\begin{equation}
  \begin{aligned}
  w_1(u) &= -L^6\cos ^6 \theta(u) \left({B_x}^2 + {g_{xx}}^2\right)  \left({E_x}^2 + g_{xx} \left(g_{tt} + g_{\Psi\Psi} \omega^2\right)\right) \left(g_{uu} + g_{\theta\theta} \theta'(u)^2\right) ,\\
  w_2(u) &= -L^6\cos ^6 \theta(u) \left({B_x}^2 + {g_{xx}}^2\right) \left(g_{tt} + g_{\Psi\Psi} \omega^2\right), \\
  w_3(u) &= -L^6\cos ^6 \theta(u) \,g_{xx} \left({E_x}^2 + g_{xx} \left(g_{tt} + g_{\Psi\Psi} \omega^2\right)\right), \\
  w_4(u) &= -L^6\cos ^6 \theta(u) \left({E_x}^2 + g_{tt} g_{xx}\right) \left({B_x}^2 + {g_{xx}}^2\right)  g_{\Psi\Psi}, \\
  w_5(u) &= -2 L^6\cos ^6 \theta(u) \, E_x \left({B_x}^2 + {g_{xx}}^2\right)  g_{\Psi\Psi} \omega, \\
  w_6(u) &= B_x \cos^4\theta(u).
  \end{aligned}
\end{equation}
Note that the Wess-Zumino term is indeed non-vanishing, indicating that the contribution of the anomaly is being incorporated.
$\mathcal{N}$ is the same as the one used in Section \ref{subsec:32}.

We find the following quantities are constant: 
\begin{align}
  \wt{C}_x &=-\frac{1}{\mathcal{N}}\frac{\partial \mathcal{L}_{\mathrm{D7}}}{\partial a_x'(u)}= -  \omega w_6(u) + \frac{2 w_2(u) a_x'(u) + w_5(u) \Phi'(u)}{2 C} ,\\
  \wt{C}_y &= -\frac{1}{\mathcal{N}}\frac{\partial \mathcal{L}_{\mathrm{D7}}}{\partial a_y'(u)}= \frac{w_3(u) a_y'(u)}{C} ,\\
  \wt{C}_z &= - \frac{1}{\mathcal{N}}\frac{\partial \mathcal{L}_{\mathrm{D7}}}{\partial a_z'(u)}= \frac{w_3(u) a_z'(u)}{C} ,\\
  \wt{C}_\phi &=- \frac{1}{\mathcal{N}}\frac{\partial \mathcal{L}_{\mathrm{D7}}}{\partial  \Phi'(u)}= - E_x w_6(u) +   \frac{w_5(u) a_x'(u) + 2 w_4(u) \Phi'(u)}{2C},
\end{align}
where 
\begin{equation}
  C=\sqrt{w_1(u) + w_2(u) a_x'(u)^2 + w_3(u) \left(a_y'(u)^2 + a_z'(u)^2\right) + w_5(u) a_x'(u) \Phi'(u) + w_4(u) \Phi'(u)^2} .\nn
\end{equation}
The current densities are read as $j_i =\mathcal{N}\wt{C}_i$ ($i=x,y,z$). 
We found that $\wt{C}_\phi$ is related to $\partial_{\mu}J_5^{\mu}$ in (\ref{eq:divj5mu}), and we identify $\partial_{\mu}J_5^{\mu}=2\mathcal{N}\wt{C}_\phi$.
The behaviors of the bulk fields near the boundary are found to be
\begin{eqnarray}
    a_x(u)&=& (\text{const.}) + \frac{\wt{C}_x-B_x\omega}{2 }u^2 + \mathcal{O} (u^3),
    \label{ax-expansion}\\
    a_y(u)&=& (\text{const.}) + \frac{\wt{C}_y}{2 }u^2 + \mathcal{O}(u^3),\\
    a_z(u)&=& (\text{const.}) + \frac{\wt{C}_z}{2 }u^2 + \mathcal{O} (u^3),\\
    \Phi (u)&=& (\text{const.}) + \frac{\wt{C}_\phi - E_x B_x}{2 }u^2 + \mathcal{O} (u^3).  \label{Phi-expansion}  
\end{eqnarray}
Note that $j_{x}$ is not directly given by the coefficient of the second term in the right-hand side of (\ref{ax-expansion}).
This is because
the Wess-Zumino term contains only $a_x^{\prime}$, whereas the DBI action does ${a_x^{\prime}}^2$. When solving the equation of motion for $a_x$, the contributions from the DBI action and the Wess-Zumino term appear in different powers of $u$. 
The same explanation applies to the second term in the right-hand side of (\ref{Phi-expansion}).

We perform the following Legendre transformation,
{\small
\begin{align}\label{eq:Llag}
\wt{\mathcal{L}}_\mathrm{D7} &= \mathcal{L}_\mathrm{D7} - \left[ a_x'(u) \frac{\partial \mathcal{L}_{\mathrm{D7}}}{\partial a_x'(u)} + a_y'(u) \frac{\partial \mathcal{L}_{\mathrm{D7}}}{\partial a_y'(u)} + a_z'(u) \frac{\partial \mathcal{L}_{\mathrm{D7}}}{\partial a_z'(u)} + \Phi'(u) \frac{\partial \mathcal{L}_{\mathrm{D7}}}{\partial \Phi'(u)} \right], \nn \\
&= \mathcal{N}\Bigg[ w_1(u) \left(1 - \frac{P^2}{w_2(u)} \right) - \frac{w_1(u) }{w_3(u)}\left(\wt{C}_y^2 + \wt{C}_z^2\right) - \frac{4 w_1(u) w_2(u)}{4 w_2(u) w_4(u) - {w_5(u)}^2} \left( Q -\frac{w_5(u)}{2 w_2(u)}P\right)^2 \Bigg]^{1/2},
\end{align}
}
where
\begin{gather}
  w_1(u)=-\left({B_x}^2 + {g_{xx}}^2\right) {(L^2 \cos ^2\theta)}^3 \left({E_x}^2 + g_{xx} \left(-|g_{tt}| + g_{\Psi\Psi} \omega^2\right)\right) \left(g_{uu} + g_{\theta\theta} \theta'(u)^2\right), \\
  \frac{w_1(u)}{w_3(u)}=\frac{\left({B_x}^2 + {g_{xx}}^2\right) \left(g_{uu} + g_{\theta\theta} \theta'(u)^2\right)}{g_{xx}} \geq 0,\\
  \frac{4 w_1(u) w_2(u) }{4 w_2(u) w_4(u) - {w_5(u)}^2} =\frac{(|g_{tt}| -\omega^2 g_{\Psi\Psi} )(g_{uu}+g_{\theta \theta} \theta ^{\prime }(u)^2)}{|g_{tt}|  g_{\Psi\Psi}},
\end{gather}
and $P=\left(\wt{C}_x - \omega w_6(u)\right),Q= \left( \wt{C}_\phi - E_x w_6(u)\right)$.

Let us find the conditions under which $\wt{\mathcal{L}}_\mathrm{D7}$ is real over the entire domain of the bulk.
We define $u=u_\ast$ as the location of $w_1(u)$ vanishes:
\begin{equation}\label{eq:effHorizon}
  \left.\left({E_x}^2 + g_{xx} \left(-|g_{tt}| + g_{\Psi\Psi} \omega^2\right)\right)  \right|_{u= u_\ast} =0.
\end{equation}
We find that
\begin{equation}
{4 w_1(u_\ast) w_2(u_\ast) }/{[4 w_2(u_\ast) w_4(u_\ast) - {w_5(u_\ast)}^2]}>0.
\label{term2}
\end{equation}
Therefore, for (\ref{eq:Llag}) to be real at $u=u_\ast$, we request
\begin{gather}
  \wt{C}_y=\wt{C}_z=0 , \label{Ctildezero}\\
  \left. \left(Q -\frac{w_5(u)}{2 w_2(u)}P  \right)\right|_{u=u_\ast}= 0 \label{eq:cond1}.
\end{gather}
We find $j_y=j_z=0$ from (\ref{Ctildezero}).

To determine the values of $\wt{C}_x$ and $\wt{C}_{\phi}$, we need one more condition in addition to (\ref{eq:cond1}). Let us find the remaining condition.
Substituting $\wt{C}_y=\wt{C}_z=0$ into (\ref{eq:Llag}), we have
\begin{align}
\wt{\mathcal{L}}_\mathrm{D7} = \mathcal{N}\Bigg[ w_1(u) \left(1 - \frac{P^2}{w_2(u)} \right) - \frac{4 w_1(u) w_2(u)}{4 w_2(u) w_4(u) - {w_5(u)}^2} \left( Q -\frac{w_5(u)}{2 w_2(u)}P\right)^2 \Bigg]^{1/2}.
\end{align}
Since the sign of $w_1$ changes from negative to positive when $u$ goes across $u=u_\ast$ from $u>u_\ast$ to $u<u_\ast$, $w_1$ is expanded as
\begin{equation}
 w_1(u) = - a (u-u_\ast) + \cdots,   
\end{equation}
with a positive constant $a$.
We also find from (\ref{eq:cond1}) and (\ref{term2}) that
\begin{gather}
  - \frac{4 w_1(u) w_2(u)}{4 w_2(u) w_4(u) - {w_5(u)}^2} \left( Q -\frac{w_5(u)}{2 w_2(u)}P\right)^2  = -b (u-u_\ast)^2 + \cdots,
\end{gather}
with a positive constant $b$.
Together with the following expansion,
\begin{equation}\label{eq:cond2}
   \left(1 - \frac{P^2}{w_2(u)} \right) = c_0 + c_1 (u-u_\ast) + c_2 (u-u_\ast)^2 + \cdots ,
\end{equation}
we obtain
\begin{equation}
  \wt{\mathcal{L}}_\mathrm{D7} =\mathcal{N}\left[ - ac_0 (u-u_\ast )-(a c_1+b) (u-u\ast)^2  
  +\cdots  \right]^{1/2}.
\end{equation}
In order to make $\wt{\mathcal{L}}_\mathrm{D7}$ real in the vicinity of $u_\ast$, we must have $c_0=0$ and $(-ac_1 -b)\geq 0$.
Using these two conditions, we can ultimately rewrite (\ref{eq:cond2}) as
\begin{equation}
  \left(1 - \frac{P^2}{w_2(u)} \right) = c_1 (u-u_\ast) + \cdots \qquad (c_1 <0),
\end{equation} 
and we obtain the following condition:
\begin{equation}\label{eq:cond3}
  \left.\left(1 - \frac{P^2}{w_2(u)} \right)\right|_{u=u\ast}=0.
\end{equation}
Using (\ref{eq:cond1}) and (\ref{eq:cond3}), we can determine $j_x  = \mathcal{N}\wt{C}_x$ and $\partial_{\mu}j_5^{\mu}  = 2\mathcal{N}\wt{C}_\phi$.
$j_x$ is explicitly given as
\begin{align}\label{eq:result_jx}
    j_x &=  \mathcal{N}(\omega w_6(u_\ast )+ \sqrt{w_2(u_\ast )}), \nn \\
  &=\mathcal{N} B_x  L^4 \omega \cos^4(\theta(u_\ast))  + \mathcal{N}E_x \sqrt{(B_x^2 + g_{xx}(u_\ast )^2)g_{\Psi \Psi }^3 (u_\ast )g_{xx}(u_\ast )^{-1}}, \nn\\
  &= \frac{N_c}{2\pi^2} \mu_5 B_x  \cos^4(\theta(u_\ast))
  +  \frac{N_c}{(2\pi)^2} \frac{E_x  \cos^3(\theta(u_\ast))}{{u_H}^2 u_\ast} \sqrt{\frac{2\pi^2 \lambda^{-1}{B_x}^2 {u_H}^8 {u_\ast}^4 +  \left({u_H}^4 + {u_\ast}^4\right)^2}{({u_H}^4 + {u_\ast}^4)}},
\end{align}
where we have used $ \mathcal{N} L^4 =N_c/(2\pi)^2$ and $\omega = 2\mu_5$ in the last line.

The first term in (\ref{eq:result_jx}) represents the current density due to the chiral magnetic effect.\footnote{The chiral magnetic effect in the D3/D7 model was discussed in \cite{Hoyos:2011us}.}
 Note that the first term in (\ref{eq:result_jx}) depends not only on $\mu_5$ but also on $E_x$ through $u_\ast$ in our setup.
When the fermions in the flavor sector are massless and $\theta(u)=0$, the first term reproduces (\ref{eq:CME}).
The second term in (\ref{eq:result_jx}) represents the current density that does not originate from the chiral anomaly, corresponding to $\bs{j}_{\mathrm{Ohm}}$ in Section \ref{sec:2}.

For $\partial_{\mu}j_5^{\mu}=2\mathcal{N}\wt{C}_\phi$, we obtain 
\begin{align}\label{eq:result_delj5}  
   \partial_{\mu}j_5^{\mu}&=  2\mathcal{N} E_x w_6(u_\ast )+\mathcal{N}\frac{w_5(u_\ast )}{\sqrt{w_2(u_\ast)}}, \nn \\
   &=  \frac{N_c}{2\pi^2} B_x E_x  \cos^4(\theta(u_\ast)) \nn \\
  & \qquad -  \frac{ \mu_5 N_c\lambda \cos^3(\theta(u_\ast)) \sin^2(\theta(u_\ast))}{4\pi^4{u_H}^6 {u_\ast}^3} \sqrt{\left({u_H}^4 + {u_\ast}^4\right) \left(2\pi^2 \lambda^{-1}{B_x}^2 {u_H}^8 {u_\ast}^4 +  \left({u_H}^4 + {u_\ast}^4\right)^2\right)}.
\end{align}
The first term in (\ref{eq:result_delj5}) represents production of the axial charge, while the second term represents their dissipation.
By imposing a steady-state condition $\del_\mu j^{\mu}_5=0$ on (\ref{eq:result_delj5}) so that the production and dissipation of the axial charge balance, we obtain
\begin{equation}\label{eq:mu5}
    \mu_5 = \frac{2\pi^2 B_x E_x u_H^6 u_\ast^3 \cos(\theta(u_\ast))}{\lambda \sin^2(\theta(u_\ast)) \sqrt{(u_H^4 + u_\ast^4) \left[ 2\pi^2 \lambda^{-1}B_x^2 u_H^8 u_\ast^4 + (u_H^4 + u_\ast^4)^2 \right]}}.
\end{equation}

Note that the right-hand side of (\ref{eq:mu5}) contains $u_\ast$, and $u_\ast$ is given by (\ref{eq:effHorizon}) as a function of $\omega=2\mu_5$. Therefore, we need to solve (\ref{eq:mu5}) and (\ref{eq:effHorizon}) simultaneously.
From (\ref{eq:mu5}) and (\ref{eq:effHorizon}), $u_{\ast}$ is given as a function of $u_{H}, B_x, E_x$ and $\theta(u_\ast)$ by eliminating $\mu_5$.
That is explicitly written as
{\small
\begin{equation}\label{eq:u_ast}
\begin{split}
u_{\ast} = \frac{u_H}{\sqrt{2}} \Biggl\{ 
& \left[ \left(4 + 2\pi^2 \lambda^{-1}(B_x^2 + E_x^2) u_H^4\right)^2 + 8\pi^4 \lambda^{-2} B_x^2 E_x^2 u_H^8 \cot^2(\theta(u_\ast)) \right]^{1/2} \\
&- u_H^2 \biggl( 2\pi^2 \lambda^{-1} (B_x^2 - E_x^2) u_H^2 \\
&+ \Bigl[ 16\pi^2 \lambda^{-1}(B_x^2 + E_x^2) + 8\pi^4 \lambda^{-2} (B_x^4 + E_x^4) u_H^4 + 8\pi^4 \lambda^{-2} B_x^2 E_x^2 u_H^4 \cot^2(\theta(u_\ast)) \\
&\quad + 4\pi^2 \lambda^{-1} (E_x^2 - B_x^2) \left( (4 + 2\pi^2 \lambda^{-1}(B_x^2 + E_x^2) u_H^4)^2 + 16 \pi^4 \lambda^{-2} B_x^2 E_x^2 u_H^8 \cot^2(\theta(u_\ast)) \right)^{1/2} \Bigr]^{1/2} \biggr) \Biggr\}^{1/4}.
\end{split}
\end{equation}
}
Substituting (\ref{eq:mu5}) into (\ref{eq:result_jx}), we obtain our final result
\begin{equation}
    j_x = \frac{N_c E_x \cos^3(\theta(u_\ast)) \left[ 2\pi^2 B_x^2 u_H^8 u_\ast^4 (1 + \cos^2(\theta(u_\ast))) + \lambda \sin^2(\theta(u_\ast)) (u_H^4 + u_\ast^4)^2 \right]}{4\pi^2 \lambda u_H^2 u_\ast \sin^2(\theta(u_\ast)) \sqrt{(u_H^4 + u_\ast^4) \left[ 2\pi^2 \lambda^{-1}B_x^2 u_H^8 u_\ast^4 + (u_H^4 + u_\ast^4)^2 \right]}}.
\end{equation}
The electric resistivity $\rho$ is read as
\begin{equation}\label{eq:rho}
    \rho = \frac{4\pi^2  u_H^2 u_\ast \sin^2(\theta(u_\ast)) \sqrt{\lambda(u_H^4 + u_\ast^4) \left[ 2\pi^2 B_x^2 u_H^8 u_\ast^4 + \lambda (u_H^4 + u_\ast^4)^2 \right]}}{N_c \cos^3(\theta(u_\ast)) \left[ 2\pi^2 B_x^2 u_H^8 u_\ast^4 (1 + \cos^2(\theta(u_\ast))) + \lambda \sin^2(\theta(u_\ast)) (u_H^4 + u_\ast^4)^2 \right]}.
\end{equation}
We find $\rho$ is a decreasing function of $B_x$ and the negative magnetoresistance is realized.
We demonstrate the behaviour of $\rho$ explicitly by numerical computations in the next subsection.


\subsection{Numerical results}\label{subsec:43}
In this subsection, we will demonstrate the behavior of the resistivity (\ref{eq:rho}).
To determine the value of $\theta(u_\ast)$ in (\ref{eq:rho}), we need to solve the equation of motion for $\theta(u)$ numerically, since the equation of motion is a nonlinear differential equation.
For numerical computations, we set 
$\lambda =2\pi^2$, and $N_c=(2\pi)^2$.\footnote{This corresponds to $L=\mathcal{N}=1$. $N_c$ should be an integer, but even if we set it to $(2\pi)^2$, the arguments of this paper remain entirely intact.}
 We further set $u_H=1$.\footnote{
In this section, the temperature of the system is fixed at $\sqrt{2}/\pi$. However, one can extend the results to other temperatures by interpreting them as if all physical quantities are nondimensionalized using $u_H$.
}
 
We adopt the shooting method.
We set $E_x$ and $B_x$ to chosen values and set $\theta(u_\ast)$ to a specific value. Then $u_\ast$ is obtained as a function of $E_x, B_x$, and $\theta (u_\ast )$ (and $u_{H}$ which we have set $u_{H}=1$ here).  
The equation of motion for $\theta(u)$ is reduced to a first-order differential equation at $u=u_\ast$, and then $\theta^{\prime}(u_\ast)$ is specified through it. 
$\theta(u_\ast)$ and $\theta^{\prime}(u_\ast)$ given here are boundary conditions for the equation of motion for $\theta(u)$, which we solve numerically. We read $m$ in (\ref{eq:mass}) with the obtained numerical solution. As a result, we obtain a solution with given $E_x, B_x$ and $m$. We repeat the above procedure while gradually changing the value of $\theta(u_\ast)$. 
By extracting the data where $m$ and $E_x$ fall within a specified tolerance for the desired values, we obtain combinations of $(B_x, j_x)$ for given $m$ and $E_x$.

\begin{figure}[htbp]
\begin{center}
\includegraphics[width=.8\textwidth]{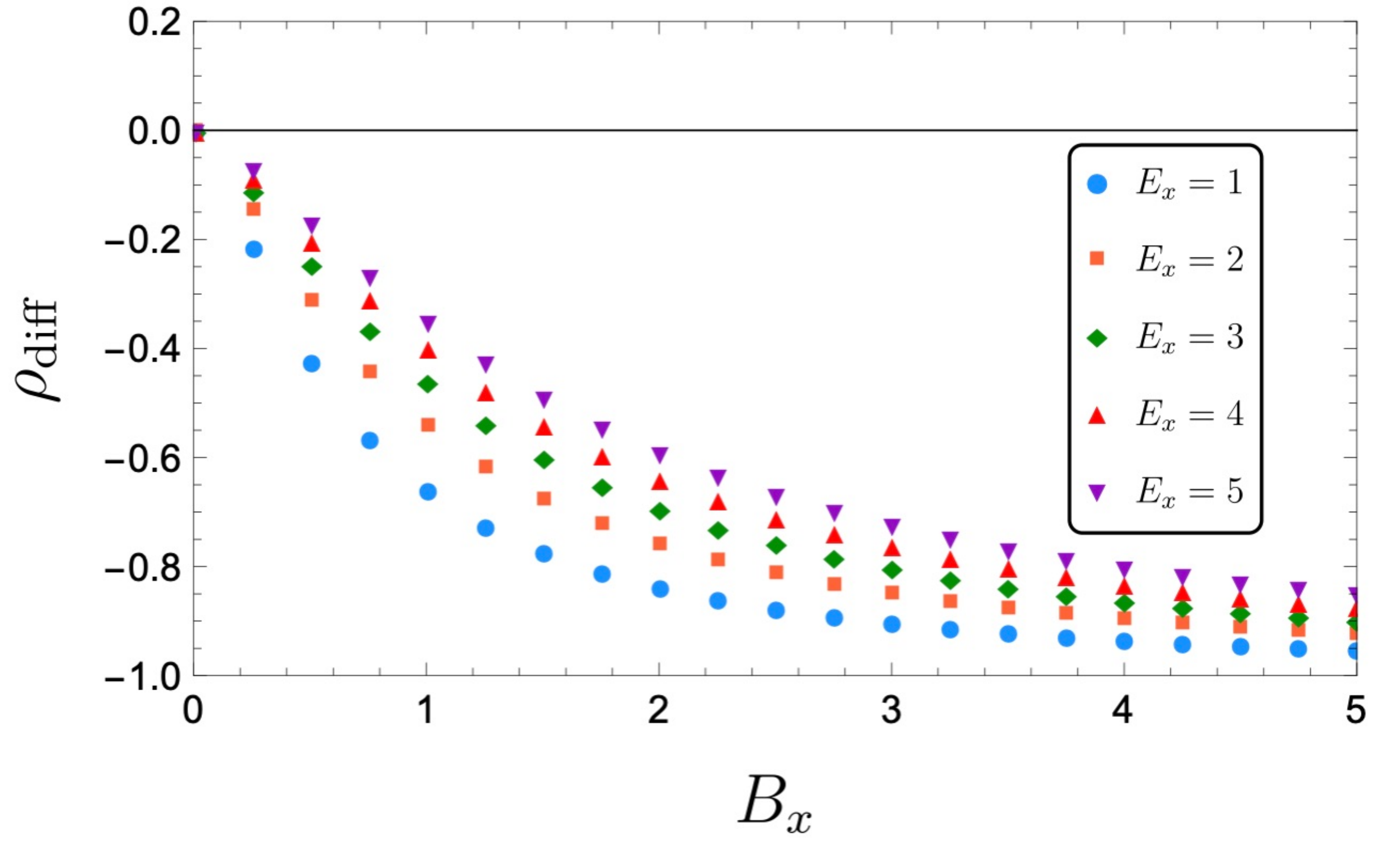}
\caption{$B_x$ dependence of $\rho_{\mathrm{diff}}$ at various values of $E_x$.}
\label{fig:PLOT1}
\end{center}
\end{figure}

Figure \ref{fig:PLOT1} shows plots of these data points, where $\rho_{\mathrm{diff}}$ is presented instead of $j_{x}$.
Here, $\rho_{\mathrm{diff}}(B_x)$ is a normalized resistivity that is defined as
\begin{equation}
  \rho_{\mathrm{diff}}= \frac{\rho (B_x)-\rho (\epsilon)}{\rho (\epsilon)},
\end{equation}
where $\rho (B_x)$ is the resistivity at given $B_x$ and $\epsilon$ is a sufficiently small cutoff that has been introduced for stability of numerical computations. $\epsilon=0.001$ in our analysis.
We have set $m=0.3$ in Figure \ref{fig:PLOT1}.

\begin{figure}[htbp]
\begin{center}
\includegraphics[width=.8\textwidth]{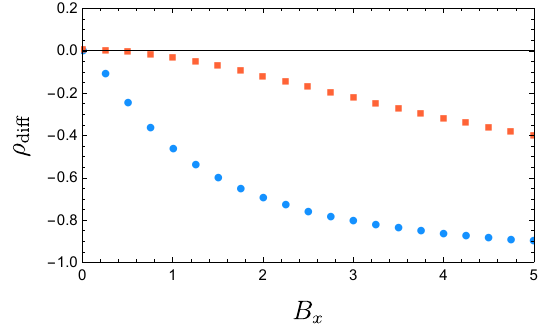}
\caption{Comparison of the $\rho _\mathrm{diff} -B_x$ characteristics between the computations with and without the contributions of the anomaly.
Blue circles represent the results of our model with contributions of the anomaly, while orange squares represent the results obtained in (\ref{eq:rhoAmmon}) where the contributions of the anomaly are absent.
}
\label{fig:PLOT2}
\end{center}
\end{figure}

It is clearly seen that electric resistance decreases as the magnetic field $B_x$ increases.
Figure \ref{fig:PLOT2} shows the magnetic-field dependence of $\rho _\mathrm{diff}$ with and without the contribution of the anomaly at $E_{x}=1$ and $m=0.3$.
The blue circles represent the results of our model with the contributions of the anomaly, while the orange squares represent the results of \cite{Ammon:2009jt} given in (\ref{eq:rhoAmmon}). 

We find that the electric resistivity decreases more rapidly when the contributions of the anomaly are taken into account.


\section{Conclusion and discussions}\label{sec:5}
We have developed a consistent framework for computing nonlinear electric conductivity that properly incorporates the effects of the chiral anomaly in the D3/D7 model.
The crucial ingredient of this framework is a D7-brane configuration rotating within the $S^5$. By employing this rotating configuration, we incorporate the axial chemical potential $\mu_5$ (or the axial charge density $n_5$) generated by the electromagnetic field through the anomaly.
The magnitude of $\mu_5$ is dynamically determined by imposing a steady-state condition. The steady state considered here is a non-equilibrium steady state realized through a balance between the production of axial charge by the anomaly and its dissipation into the thermal bath.
Our numerical results show that the anomaly contribution enhances the magnitude of the negative magnetoresistance.

It is interesting to investigate the dissipation mechanisms in our system further. In our system, the momentum of charge carriers, energy, and axial charge dissipate into the thermal bath. It is discussed in \cite{Amoretti:2023hpb} that a generalized dissipation that couples these three channels is necessary, and the system must be an open system to realize electric charge conservation, Onsager reciprocity, and finite DC conductivity. While it is clear that our system is an open system in contact with a thermal bath, it would be beneficial to further investigate the structure of the relaxation process in our system.

The calculation scheme presented here has various applications.
For example, we can apply our scheme to the model of holographic Weyl semimetals based on the D3/D7 model \cite{BitaghsirFadafan:2020lkh}.
Since the negative magnetoresistance is one of the most intriguing phenomena associated with Weyl semimetals, computing the magnetoresistance in holographic Weyl semimetals is a crucial application.

Other applications include studies of non-equilibrium steady states. 
Phase transitions of non-equilibrium steady states driven by a finite current density along the external electric field have been studied in holography \cite{Nakamura:2012ae,Ali-Akbari:2013hba,Matsumoto:2018ukk,Vahedi:2018gvn,Imaizumi:2019byu,Matsumoto:2022nqu,Endo:2023uqk}. Phase transitions of a system moving at finite velocity with respect to the thermal bath were also studied in holography \cite{Nakamura:2025grh} from the perspective of proper effective temperature \cite{Hoshino:2018vne}.
In the present setup, the flavor sector is in a non-equilibrium steady state driven by both a finite current and axial charge production by the anomaly.
Our work provides a framework for studying systems in such non-equilibrium steady states.

\begin{acknowledgments}
The authors acknowledge M. Matsumoto and N. Yamamoto for discussions and comments.
The work of S.\,N.~is supported in part by JSPS KAKENHI Grant No.~JP25K07174, and the Chuo University Personal Research Grant.
\end{acknowledgments}



 \bibliographystyle{JHEP}
 \bibliography{biblio.bib}

\end{document}